# A Unified Subspace Outlier Ensemble Framework for Outlier Detection in High Dimensional Spaces[*]

Zengyou He, Xiaofei Xu, Shengchun Deng

*Department of Computer Science and Engineering Harbin Institute of Technology,*

*92 West Dazhi Street, P.O Box 315, P. R. China, 150001*

Email: zengyouhe@yahoo.com, {xiaofei, dsc}@hit.edu.cn

**Abstract** The task of outlier detection is to find small groups of data objects that are exceptional when compared with rest large amount of data. Detection of such outliers is important for many applications such as fraud detection and customer migration. Most such applications are high dimensional domains in which the data may contain hundreds of dimensions. However, the outlier detection problem itself is not well defined and none of the existing definitions are widely accepted, especially in high dimensional space.

In this paper, our first contribution is to propose a unified framework for outlier detection in high dimensional spaces from an ensemble-learning viewpoint. In our new framework, the outlying-ness of each data object is measured by fusing outlier factors in different subspaces using a combination function. Accordingly, we show that all existing researches on outlier detection can be regarded as special cases in the unified framework with respect to the set of subspaces considered and the type of combination function used.

In addition, to demonstrate the usefulness of the ensemble-learning based outlier detection framework, we developed a very simple and fast algorithm, namely SOE1 (**S**ubspace **O**utlier **E**nsemble using **1**-dimensional Subspaces) in which only subspaces with one dimension is used for mining outliers from large categorical datasets. The SOE1 algorithm needs only two scans over the dataset and hence is very appealing in real data mining applications. Experimental results on real datasets and large synthetic datasets show that: (1) SOE1 has comparable performance with respect to those state-of-art outlier detection algorithms on identifying true outliers and (2) SOE1 can be an order of magnitude faster than one of the fastest outlier detection algorithms known so far.

# 1 Introduction

In contrast to traditional data mining task that aims to find the general pattern applicable to the majority of data, outlier detection targets the finding of the rare data whose behavior is very exceptional when compared with rest large amount of data. Studying the extraordinary behavior of outliers helps uncovering the valuable knowledge hidden behind them and aiding the decision makers to make profit or improve the service quality. Thus, mining for outliers is an important data mining research with numerous applications, including credit card fraud detection, discovery of criminal activities in electronic commerce, weather prediction, and marketing.

---

[*] This work was supported by the High Technology Research and Development Program of China (No. 2002AA413310, No. 2003AA4Z2170, 2003AA413021) and the IBM SUR Research Fund.

A well-quoted definition of outliers is firstly given by Hawkins[1]. This definition states, "An outlier is an observation that deviates so much from other observations as to arouse suspicion that it was generated by a different mechanism". With increasing awareness on outlier detection in data mining literature, more concrete meanings of outliers are defined for solving problems in specific domains [3-37]. Nonetheless, most of these definitions follow the spirit of the Hawkins-Outlier.

Most applications for outlier mining are high dimensional domains in which the data may contain hundreds of dimensions. However, most existing techniques [3-37] try to define outliers by using the full dimensional distances of the points from one another. Recent research results show that in high dimensional space, the concept of proximity may not be qualitatively meaningful [2]. Due to the *curse of dimensionality*, most existing approaches are not appropriate for discovering outliers in high dimensional space.

To overcome *curse of dimensionality*, Aggarwal and Yu [3] proposed a definition for outliers in low-dimensional projections and developed an evolutionary algorithm for finding outliers in projected subspaces.

A frequent pattern based outlier detection method is proposed in [4], which aims at utilizing frequent patterns in different subspaces together to define outliers in high dimensional space.

Overall, although there are so many definitions for outliers, we can see that none of the existing definitions are widely accepted and the outlier detection problem in high dimensional spaces is also not well defined. One natural question one may ask is: "*Can we formally define the problem of mining outliers in high dimensional spaces and unify existing researches in a more general framework?*" To answer this question, in this paper, we make a first step towards formally introducing the problem and proposing a unified framework for outlier detection in high dimensional spaces from an ensemble-learning viewpoint.

In our new framework, the outlying-ness of each data object is measured by fusing outlier factors in different subspaces using a combination function. Accordingly, we show that all existing researches on outlier detection can be regarded as special cases in the unified framework with respect to the set of subspaces considered and the type of combination function used.

In addition, to demonstrate the usefulness of the ensemble-learning based outlier detection framework, we developed a very simple and fast algorithm, namely SOE1 (**S**ubspace **O**utlier **E**nsemble using **1**-dimensional Subspaces) in which only subspaces with one dimension is used for mining outliers from large categorical datasets. The SOE1 algorithm needs only two scans over the dataset and hence is very appealing in real data mining applications. Experimental results on real datasets and large synthetic datasets show that: (1) SOE1 has comparable performance with respect to those state-of-art outlier detection algorithms on identifying true outliers and (2) SOE1 can be an order of magnitude faster than one of the fastest outlier detection algorithms known so far.

The organization of this paper is as follows. First, we review related work in the next section. Problem formulation and the proposed unified framework are provided in Section 3. Sections 4 presents the SOE1 algorithm. Empirical studies are provided in Section 5 and a section of concluding remarks follows.

# 2 Related Work

Previous researches on outlier detection broadly fall into the following categories.

*Distribution based* methods are in the first category, which are previously conducted by the statistics community [1,5,6]. They deploy some standard distribution model (e.g., normal) and flag as outliers those points that deviate from the model. Recently, Yamanishi, Takeuchi and Williams [7] used a Gaussian mixture model to present the normal behaviors and each datum is given a score based on changes in the model. High score indicates high possibility of being an outlier. This approach has been combined with a supervised-based learning approach to obtain general patterns for outlier [8]. For arbitrary data sets without any prior knowledge of the distribution of points, we have to perform expensive tests to determine which model fits the data best, if any.

*Depth-based* is the second category for outlier mining in statistics [9,10]. Based on some definition of depth, data objects are organized in convex hull layers in data space according to peeling depth, and outliers are expected to be detected out from data objects with shallow depth values.

*Deviation-based* techniques identify outliers by inspecting the characteristics of objects and consider an object that deviates these features as an outlier [11].

*Distance based* method was originally proposed by E.M. Knorr and R.T. Ng [12-15]. A distance-based outlier in a dataset $D$ is a data object with $pct\%$ of the objects in $D$ having a distance of more than $d_{min}$ away from it. This notion generalizes many concepts from distribution-based approach and enjoys better computational complexity. It is further extended based on the distance of a point from its $k^{th}$ nearest neighbor [16]. After ranking points by the distance to its $k^{th}$ nearest neighbor, the *top k* points are identified as outliers. Efficient algorithms for mining *top-k* outliers are given. Alternatively, in the algorithm proposed by Angiulli and Pizzuti [17], the outlier factor of each data point is computed as the sum of distances from its $k$ nearest neighbors. Bay and Schwabacher [18] modify a simple algorithm based on nested loops to yield near linear time mining for distance-based outlier detection.

*Density based* This was proposed by M. Breunig, et al. [19]. It relies on the local outlier factor *(LOF)* of each point, which depends on the local density of its neighborhood. In typical use, points with a high *LOF* are flagged as outliers. Tang el at [20] introduces a connectivity-based outlier factor (*COF*) scheme that improves the effectiveness of *LOF* scheme when a pattern itself has similar neighborhood density as an outlier. Three enhancement schemes over *LOF* are introduced in [21]. An effective algorithm for mining local outliers is proposed in [22]. The *LOCI* method [23] and low-density regularity method [24] further extended the density-based approach [19].

*Clustering-based* outlier detection techniques regarded *small* clusters as outliers [25] or identified outliers by removing clusters from the original dataset [26]. The authors in [27] further extended existing clustering based techniques by proposing the concept of *cluster-based local outlier*, in which a measure for identifying the outlier-ness of each data object is defined.

*Sub-Space based.* Multi- and high-dimensional data make the outlier mining problem more complex because of the impact of *curse of dimensionality* on algorithms' both performance and effectiveness. Aggarwal and Yu [3] discussed a new technique for outlier detection, which finds outliers by observing the density distribution of projections from the data. A frequent pattern based outlier detection method is proposed in [4], which aims at utilizing frequent patterns in different subspaces together to define outliers in high dimensional space. Wei, et al. [28] introduces an outlier mining method based on a hypergraph model to detect outliers in categorical dataset.

*Support vector based.* Support vector novelty detector (*SVND*) was recently developed. The first *SVND* is proposed by Tax and Duin [29], which estimates a sphere to contain all the normal data patterns with the smallest radius; the outliers can be identified from the normal data patterns by calculating the distance of a data pattern to the center of the sphere. Another alternative *SVND* is proposed by Schölkopf et al. [30]. Instead of a sphere, a function is estimated to separate the region of normal data patterns from that of outliers with maximal margin, thus detecting the outliers from the normal data patterns. Cao et al. [31] further extended the *SVND* method. Petrovskiy [32] combine kernel methods and fuzzy clustering methods.

*Neutral network based.* The replicator neutral network (*RNN*) is employed to detect outliers by Harkins, et al. [33,34]. The approach is based on the observation that the trained neutral network will reconstruct some small number of individuals poorly, and these individuals can be considered as outliers. The outlier factor for ranking data is measured according to the magnitude of the reconstruction error.

In addition, the class outlier detection problem is considered in [35-37].

## 3 Problem Formulation and Unified Framework

In this section, we formalize the problem of outlier detection in high dimensional space and propose a unified framework from an ensemble-learning viewpoint.

Let $D$ be a database of $d$-dimensional feature vectors. An element $P \in D$ is called point or object. Let $A = \{A_1, A_2, ..., A_d\}$ be the set of all attributes $A_i$ of $D$. Any subset $S \subseteq A$, is called a subspace. The cardinality of $S$ ($|S|$) is called the dimensionality of $S$. We let the power set of $A$, denoted by $Pow(A)$, be the set of all unions of subsets of $A$. That is, we define $Pow(A) = \{S \mid S \subseteq A\}$. Hence, each subspace is an element of $Pow(A)$. The projection of an object $P$ into a subspace $S \in Pow(A)$ is denoted by $\pi_S(P)$. The outlier factor of an object $P$ in subspace $S$ is denoted by $OF(\pi_S(P))$.

We further assume that the outlier factor values are normalized onto the interval [0,1]. For each object's *degree* of being an outlier is measured by its outlier factor, without loss of generality, it is assumed that the higher is the outlier factor value of an object, the higher of its outlier-ness is.

The problem of outlier detection in high dimensional space and the unified ensemble learning based algorithmic framework are described in Fig.1.

The input for outlier detection in high dimensional space includes the target database, the number of desired outliers, the set of subspaces considered in the mining process and the combination function. Among all these input parameters, the set of subspaces and combination function are of primary importance. As we can see later, existing researches on outlier detection can be distinguished when they are set to different values.

**Outlier Detection in High Dimensional Space:**
*Mining top-k outliers with a set of subspaces and a combination function in a database*

**Input:**
 (1): A database $D$ with set of features $A$
 (2): An Integer $k$, i.e., the $k$ most outlying objects to be mined
 (3): $SS$, a set of subspaces, i.e., $SS$ is a subset of $Pow\ (A)$
 (4): A combination/ensemble function $\oplus$

**Output:**
  Top-k outliers that satisfy the requirement

**Unified Algorithmic Framework:**
 (1) **Individual subspace outlier factor computation step**
     For each subspace $S$ in $SS$
         For each object $P$ in $D$
             Compute the outlier factor of $P$ in $S$, i.e., $OF(\pi_S(P))$
 (2) **Outlier ensemble step**
     For each object $P$ in $D$
     Ensemble all the outlier factors of $P$ in different subspaces, i.e., $OF(P) = \underset{S \in SS}{\oplus} OF(\pi_S(P))$

**Fig. 1** The unified outlier ensemble based algorithmic framework-SOE framework

The unified algorithmic framework (SOE framework) is a two-step mining strategy, which consists of two steps: subspace outlier mining step and subspace outlier ensemble step.

# A subspace outlier mining step

In the subspace outlier-mining step, the SOE framework uses existing outlier mining algorithms to compute the outlier factors for each data objects in all the input subspaces.

  *Remarks*

(1) In different subspaces, we can choose different outlier detection algorithms to compute outlier factors for each data object. Such flexibility is very useful in real data mining applications, since different subspaces may have significant differences on attribute types and data distributions. For example, if the input database is mixed with categorical attributes and numeric attributes, we can just choose those well-designed algorithms for categorical data, such as algorithms of [4, 28], to perform outlier factor computation in pure categorical subspace. Similarly, density based outlier detection algorithms [19] will be good choice for numeric subspace.

(2) In some applications, the user maybe is only interested in the outliers detected in some specific subspace. In such situation, it is obviously that we can just finish the mining process without proceeding to the outlier ensemble stage. Furthermore, the output results can also be regarded as running our algorithmic framework for multiple times, and each run is fed with *only one* subspace.

(3) The set of input subspaces can incorporate the business user's expectation or interestingness. For example, the business user can only specify those subspaces that are meaningful in real business as input. It also exhibits the spirit of so-called constraint data mining. Furthermore, in our current study and also previous researches, how to select only meaningful (at least meaningful from a statistical viewpoint) subspaces as input remains unaddressed. The general problem is open and provides promising future research directions.

# A subspace outlier ensemble step

After computing the outlier factors for each data object in all the input subspaces, the question then becomes, "*how can we integrate these subspace outlier mining results to get final output?*" We borrow some ideas from ensemble learning by fusing outlier factors in different subspaces using a combination function. Hence, the choice of combination function (or combining operator) is at the core of the outlier ensemble stage.

Suppose the outlier factors of an object $P$ in $D$ in different subspaces are denoted as $v_1, v_2, ..., v_m$ (the number of input subspaces is $m$). And the combining operator is denoted as $\oplus$. Hence, the outlier factor of $P$ after fusing all the subspace outlier factors is $OF(P) = \oplus(v_1, v_2, ..., v_m)$. Note that if $m=1$, then we let $\oplus(v_1, v_2, ..., v_m) = v_1$.

In the sequel, we will briefly discuss 4 operators considered in our study. Although other more complex combining operators can be designed, it is not the emphasis of this paper. Moreover, as we will show, these simple operators are enough in unifying existing researches in our SOE framework.

**Choosing a combining operator**: Our potential choices for $\oplus$ are the followings. We offer some additional insights on these choices in Section 5.

- The product operator $\prod$: $\oplus(v_1, v_2, ..., v_m) = v_1 v_2 ... v_m$.
- The addition operator $+$: $\oplus(v_1, v_2, ..., v_m) = v_1 + v_2 + ... + v_m$.
- A generalization of addition operator-it is called the $S_q$ combining rule, where $q$ is an odd natural number. $S_q(v_1, v_2, ..., v_m) = (v_1^q + v_2^q + ... + v_m^q)^{(1/q)}$. Note that the addition is simply the $S_1$ rule.
- A "limiting" version of $S_q$ rules, denoted as $S_\infty$. $S_\infty(v_1, v_2, ..., v_m)$ is defined to be equal to $v_i$, where $v_i$ has the largest absolute value among $(v_1, v_2, ..., v_m)$.

Thus, the $S_1$ combining rule is *linear* with respect to the component outlier factors. The $\prod$ and $S_q$ rules for $q>1$, on the other hand, involve a *non-linear* term for each individual outlier factor. $S_\infty$ is an especially appealing rule, since it is non-linear, particularly fast to compute, and also appears to have certain useful "sum-like" properties.

Just as we have argued, all existing researches on outlier detection can be regarded as special cases in the unified framework with respect to the set of subspaces considered and the type of combination function used. Table 1 depicts our classification on existing researches for outlier detection.

**Table 1** Classification on existing researches in our framework

| Combining operator<br>Number of Input subspaces | $\prod$ | $+$ | $S_q$ | $S_\infty$ |
|---|---|---|---|---|
| One(The subspace composed of all dimensions) | All other outlier detection researches except [3,4] | | | |
| Multiple | | [4] | | [3] |

From Table 1, we can see that most researches except for [3] and [4] define the outliers using the full dimensional distances of the points from one another and hence the number of input subspaces is only one (The subspace composed of all dimensions). Since if $m=1$, let $\oplus$ ($v_1, v_2, ..., v_m$) = $v_1$. So, it would be always right to classify these researches to anyone of the given combining operators.

Aggarwal and Yu [3] consider data objects with best (smallest) sparse coefficients in projections as outliers; hence, they use the $S_\infty$ operator in our framework. He et.al [4] use the sum of supports of all frequent patterns as outlying-ness, so it is the "+" operator. Moreover, both methods in [3] and [4] take multiple subspaces as input.

So far, we have presented the unified framework for outlier detection in high dimensional spaces from an ensemble-learning viewpoint, in which the outlying-ness of each data object is measured by fusing outlier factors in different subspaces using a combination function. We also show that all existing researches on outlier detection can be regarded as special cases in the unified framework with respect to the number of subspaces considered and the type of combination function used. In the next section, we will describe the SOE1 algorithm, which is developed according to the subspace outlier ensemble framework.

## 4 SOE1 Algorithm

To demonstrate the usefulness of the ensemble-learning based outlier detection framework, we developed a very simple and fast algorithm, namely SOE1 (**S**ubspace **O**utlier **E**nsemble using **1**-dimensional Subspaces) in which only subspaces with one dimension is used for mining outliers from large categorical datasets. In SOE1, the rationale behind using only subspaces with one dimension is as follows:

(1) One-dimensional subspace is of primary importance in most real applications and is the basic element of multiple dimensional subspaces.

(2) Since data mining algorithms have to handle very large databases, hence, using only one-dimensional subspaces will greatly improve the performance of algorithms.

(3) From the framework viewpoint, if even the simpler SOE1 algorithm is very effective in practice, we will be confident on the usefulness of the ensemble-learning based outlier detection framework.

Before introducing the SOE1 algorithms, we firstly present some additional notations used in SOE1. Let $D$ be a database of $d$-dimensional feature vectors. Let $A$ = {$A_1, A_2, ..., A_d$} be the set of all attributes $A_i$ of $D$. The value set $V_i$ is set of values of $A_i$ that are present in $D$. For each $v \in V_i$, the frequency $f$ (v), denoted as $f_v$, is number of objects $P \in D$ with $P. A_i= v$. suppose the number of distinct attribute values of $A_i$ is $p_i$. We define the histogram of $A_i$ as the set of pairs: $h_i$ = {($v_1$, $f_1$), ($v_2$, $f_2$),..., ($v_{p_i}, f_{p_i}$)}. Each element of $h_i$ is called an entry in the histogram or just a histogram entry. The histogram of the dataset $D$ is defined as: $H$ = {$h_1, h_2, ..., h_d$}.

The SOE1 algorithm needs only two scans over the dataset and hence is very appealing in real data mining applications. Moreover, the SOE1 algorithm is parameter-free, i.e., it doesn't require *any* addition parameters before the mining process.

The first scan of SOE1 is the subspace outlier-mining step, in which we construct the histogram of the dataset *D*. Intuitively, in one-dimensional space the outlying-ness of an object is determined by the occurrences of its corresponding attribute value. Hence, the outlier factor of each object $P \in D$ in subspace $A_i$ is the frequency $f(P. A_i)$. To store the histogram of the dataset *D*, we need *d* hash tables as our basic data structures (each hash table for one histogram of $A_i$). Actually, each hash table is the materialization of a histogram. Hence, we will use histogram and hash table interchangeably in the remaining parts of the paper.

The second scan of SOE1 is the subspace outlier-ensemble step, in which we fuse the outlier factors in different one-dimensional subspaces using a combination function. The possible types of combining functions have been discussed in Section 3. That is, for each object $P \in D$, we retrieve its frequencies of attribute values, i.e., outlier factors, from hash tables efficiently. Then, fusing these outlier factors to get final outlying-ness. As for reporting the top-*k* outliers, we maintain a *k*-length array for this purpose.

Now we summarize the entire mining process and present the SOE1 outlier mining algorithm in Fig.2.

---

**SOE Algorithm**

**Input:**
 (1): A database *D* with set of features *A*
 (2): An Integer *k*, i.e., the *k* most outlying objects to be mined
 (3): *SS*, a set of subspaces, i.e., *SS = A*
 (4): A combination/ensemble function $\oplus$
**Output:**
  Top-k outliers that satisfy the requirement

**Method**
 **(1) Individual subspace outlier factor computation step**
   Initialize the *d* hash tables
   For each object *P* in *D*
     Update the frequencies of each entry in the hash table
 **(2) Outlier ensemble step**
   For each object *P* in *D*
    Ensemble all the outlier factors of *P* retrieved from hash tables

---

**Fig. 2** The SOE1 Algorithm

# Time and Space Complexities

**Worst-case analysis:** The time and space complexities of the SOE1 algorithm depend on the size of dataset (*n*), the number of attributes (*d*), and the size of every histogram.

To simplify the analysis, we will assume that every attribute has the same number of distinct attributes values, *p*. Then, the algorithm has time complexity $O(n*d*p)$ in worst case.

The algorithm only needs to store *d* histograms in main memory, so the space complexity of our algorithm is $O(d*p)$.

**Practical analysis:** Categorical attributes usually have *small* domains. Typical categorical attributes domains that consist of less than a hundred or, rarely, a thousand attribute values. An important of implication of the compactness of categorical domains is that the parameter, *p*, can be regarded to be very small. And the use of hashing technique in histograms also reduces the impact of *p*. So, in practice, the time complexity of SOE1 can be expected to be $O(n*d)$.

The above analysis shows that the time complexity of SOE1 is linear to the size of dataset and the number of attributes, which makes this algorithm deserve good scalability.

## Enhancement for Real Applications

The data sets in real-life applications are usually complex. They have not only categorical data but also numeric data. Sometimes, they are *incomplete*. In this section, we discuss the techniques for handling data with these characteristics in SOE1.

**Handling numeric data.** To process numeric data, we apply the widely used binning technique [38] and choose equal-width method for its feasibility in producing varied frequency values.

**Handling missing values** To handle incomplete data, we provide two choices. In the first choice, missing values in an incomplete object will not be considered when updating histograms. In the second choice, missing values are treated as special categorical attribute values. In our current implementation, we use the second choice.

# 5 Experimental Results

A comprehensive performance study has been conducted to evaluate our SOE1 algorithm. In this section, we describe those experiments and their results. We ran our algorithm on real-life datasets obtained from the UCI Machine Learning Repository [39] to test its performance against other algorithms on identifying true outliers in both medium and high dimensional spaces. In addition, some large synthetic datasets are used to demonstrate the scalability of our algorithm.

## 5.1 Experiment Design and Evaluation Method

We used three real life datasets (*lymphography*, *cancer* and *arrhythmia*) to demonstrate the effectiveness of our algorithm against *FindFPOF* algorithm [4], *FindCBLOF* algorithm [27] and *KNN* algorithm [16]. In addition, on the *cancer* dataset, we add the results of *RNN* based outlier detection algorithm [33,34] that are reported in [33,34] for comparison, although we didn't implement the *RNN* based outlier detection algorithm. In a similar way, on the *arrhythmia* dataset, we also add the results of the algorithm developed by Aggarwal and Yu [3] for comparison.

For all the experiments, the two parameters needed by *FindCBLOF* [27] algorithm are set to 90% and 5 separately as done in [27]. For the *KNN* algorithm [16], the results were obtained using the *5-nearest-neighbour*; For *FindFPOF* algorithm [4], the parameter *mini-support* for mining frequent patterns is fixed to 10%, and the maximal number of items in an itemset is set to 5. Since the SOE1 algorithm is parameter-free, we don't need to set any parameters. Furthermore, we empirically study the impact of different combining operators on SOE1. That is, in the experiments, we report the results of SOE1 with different combining operators. For $S_q$ operator, we set *q* to 2, 5 and 7 separately.

As pointed out by Aggarwal and Yu [3], one way to test how well the outlier detection algorithm worked is to run the method on the dataset and test the percentage of points which belong to the rare classes. If outlier detection works well, it is expected that the rare classes would be over-represented in the set of points found. These kinds of classes are also interesting from a practical perspective.

Since we know the true class of each object in the test dataset, we define objects in small classes as rare cases. The number of rare cases identified is utilized as the assessment basis for comparing our algorithm with other algorithms.

## 5.2 Results on Lymphography Data

The first dataset used is the Lymphography data set, which has 148 instances with 18 attributes. The data set contains a total of 4 classes. Classes 2 and 3 have the largest number of instances. The remained classes are regarded as rare class labels for they are small in size. The corresponding class distribution is illustrated in Table 2.

Table 2. Class Distribution of Lymphography Data Set

| Case | Class codes | Percentage of instances |
|---|---|---|
| Commonly Occurring Classes | 2, 3 | 95.9% |
| Rare Classes | 1, 4 | 4.1% |

Table 3 shows the results produced by different algorithms. Here, the *top ratio* is ratio of the number of records specified as *top-k* outliers to that of the records in the dataset. The *coverage* is ratio of the number of detected rare classes to that of the rare classes in the dataset. For example, we let SOE1 (+) algorithm find the *top 16* outliers with the top ratio of 11%. By examining these 16 points, we found that 6 of them belonged to the rare classes.

Table 3: Detected Rare Classes in Lymphography Dataset

| Top Ratio (Number of Records) | Number of Rare Classes Included (Coverage) | | | | | | | | |
|---|---|---|---|---|---|---|---|---|---|
| | SOE1 ($\prod$) | SOE1 (+) | SOE1($S_q$) | | | SOE1 ($S_\infty$) | Find FPOF | Find CBLOF | KNN |
| | | | q=2 | q=5 | q=7 | | | | |
| 5% (7) | **6(100%)** | 5(83%) | 4 (67%) | 4 (67%) | 4 (67%) | 2 (33%) | 5(83%) | 4 (67%) | 4 (67%) |
| 10%(15) | 6(100%) | 6(100%) | 5(83%) | 4 (67%) | 4 (67%) | 6(100%) | 5(83%) | 4 (67%) | 6(100%) |
| 11%(16) | 6(100%) | 6(100%) | 5(83%) | 4 (67%) | 4 (67%) | 6(100%) | 6(100%) | 4 (67%) | 6(100%) |
| 15%(22) | 6(100%) | 6(100%) | 5(83%) | 5(83%) | 4 (67%) | 6(100%) | 6 (100%) | 4 (67%) | 6(100%) |
| 20%(30) | 6(100%) | 6(100%) | 6(100%) | 5(83%) | 4 (67%) | 6(100%) | 6 (100%) | 6 (100%) | 6(100%) |

One important observation from Table 3 was that, among all the potential choices of $\oplus$ we are considered in SOE1, the "+" operator and "$\prod$" operator are the clear winners in this experiment. That is, SOE1 with the "+" operator and "$\prod$" operator outperform $S_q$ and $S_\infty$ in all cases. This observation suggests that the "+" operator and "$\prod$" operator will be a better choices in practice for users. Consequent experiments also get similar conclusions. Moreover, with increase of $q$ in the $S_q$ operator, the performance of SOE1 will deteriorate.

Furthermore, in this experiment, the SOE1 algorithm with "$\prod$" operator performed the best for all cases and can find all the records in rare classes when the *top ratio* reached 5%. In contrast, for the *FindFPOF* algorithm, it achieved this goal with the *top ratio* at 10%, which is almost the twice for that of our algorithm.

### 5.3 Results on Wisconsin Breast Cancer Data

The second dataset used is the Wisconsin breast cancer data set, which has 699 instances with 9 attributes, in this experiment, all attributes are considered as categorical. Each record is labeled as *benign* (458 or 65.5%) or *malignant* (241 or 34.5%). We follow the experimental technique of Harkins, et al. [33,34] by removing some of the *malignant* records to form a very unbalanced distribution; the resultant dataset had 39 (8%) *malignant* records and 444 (92%) *benign* records[1]. The corresponding class distribution is illustrated in Table 4.

**Table 4.** Class Distribution of Wisconsin breast cancer data set

| Case | Class codes | Percentage of instances |
|---|---|---|
| **Commonly Occurring Classes** | 1 | 92% |
| **Rare Classes** | 2 | 8% |

For this dataset, we also consider the *RNN* based outlier detection algorithm [33]. The results of *RNN* based outlier detection algorithm on this dataset are reproduced from [33].

Table 5 shows the results produced by the different algorithms. Clearly, SOE1 with the "+" operator and "$\prod$"operator also outperform $S_q$ and $S_\infty$ in all cases in this dataset.

One important observation from Table 5 was that, among all of these algorithms, *RNN* performed the worst in most cases. In comparison to other algorithms, SOE1 (with "+" operator and "$\prod$"operator) achieves roughly the same average performance with respect to the number of outliers identified.

### 5.4 Results on Arrhythmia Data

Both Lymphography data and Cancer data are datasets with only medium dimensional spaces. In this section, we report our experimental results on the Arrhythmia data, which has 279 attributes corresponding to different measurements of physical and heartbeat characteristics that are used in order to diagnose arrhythmia. The dataset contains a total of 13 non-empty classes. Class 1 was the largest and corresponds to people who don't have any kind of heart disease. The remaining classes correspond to people with diseases one form or another.

As suggested in [3], we also considered those kinds of class labels that occurred less than 5% of the dataset as rare classes. The corresponding class is illustrated in Table 6.

Since most attributes in this dataset are continuous, hence, we first perform a grid discretization of the data. Each attribute is divided into 2 equal-width bins.

In this dataset, we let each algorithm report top 85 outliers, as done in [3]. Among these reported data objects, we examine how many of them belong to rare classes. Table 7 shows the results produced by the different algorithms.

---

[1] The resultant dataset is public available at: http://research.cmis.csiro.au/rohanb/outliers/breast-cancer/

Table 5: Detected Malignant Records in Wisconsin Breast Cancer Dataset

| Top Ratio (Number of Records) | Number of Rare Classes Included (Coverage) | | | | | | | | | |
|---|---|---|---|---|---|---|---|---|---|---|
| | SOE1 ($\prod$) | SOE1 (+) | SOE1($S_q$) | | | SOE1 ($S_\infty$) | Find FPOF | Find CBLOF | RNN | KNN |
| | | | q=2 | q=5 | q=7 | | | | | |
| 1%(4) | 4 (10.26%) | 4 (10.26%) | 3 (7.69%) | 3 (7.69%) | 3 (7.69%) | 3 (7.69%) | 3 (7.69%) | 4 (10.26%) | 3 (7.69%) | 4 (10.26%) |
| 2%(8) | 7 (17.95%) | 7 (17.95%) | 7 (17.95%) | 7 (17.95%) | 7 (17.95%) | 5 (12.82%) | 7 (17.95%) | 7 (17.95%) | 6 (15.38%) | 8 (20.52%) |
| 4%(16) | 15 (38.46%) | 14 (35.90%) | 14 (35.90%) | 14 (35.90%) | 14 (35.90%) | 11 (28.21%) | 14 (35.90%) | 14 (35.90%) | 11 (28.21%) | 16 (41.03%) |
| 6%(24) | 22 (56.41%) | 21 (53.85%) | 19 (48.72%) | 19 (48.72%) | 16 (41.03%) | 17 (43.59%) | 21 (53.85%) | 21 (53.85%) | 18 (46.15%) | 20 (51.28%) |
| 8%(32) | 27 (69.23%) | 28 (71.79%) | 26 (66.67%) | 25 (64.10%) | 23 (58.97%) | 23 (58.97%) | 28 (71.79%) | 27 (69.23%) | 25 (64.10%) | 27 (69.23%) |
| 10%(40) | 33 (84.62%) | 32 (82.05%) | 31 (79.49%) | 30 (76.92%) | 28 (71.79%) | 28 (71.79%) | 31 (79.49%) | 32 (82.05%) | 30 (76.92%) | 32 (82.05%) |
| 12%(48) | 36 (92.31%) | 36 (92.31%) | 34 (87.18%) | 33 (84.62%) | 33 (84.62%) | 33 (84.62%) | 35 (89.74%) | 35 (89.74%) | 35 (89.74%) | 37 (94.87%) |
| 14%(56) | 39 (100%) | 39 (100%) | 38 (97.44%) | 37 (94.87%) | 37 (94.87%) | 37 (94.87%) | 39 (100%) | 38 (97.44%) | 36 (92.31%) | 39 (100%) |
| 16%(64) | 39 (100%) | 39 (100%) | 39 (100%) | 38 (97.44%) | 38 (97.44%) | 38 (97.44%) | 39 (100%) | 39 (100%) | 36 (92.31%) | 39 (100%) |
| 18%(72) | 39 (100%) | 39 (100%) | 39 (100%) | 39 (100%) | 39 (100%) | 38 (97.44%) | 39 (100%) | 39 (100%) | 38 (97.44%) | 39 (100%) |
| 20%(80) | 39 (100%) | 39 (100%) | 39 (100%) | 39 (100%) | 39 (100%) | 39 (100%) | 39 (100%) | 39 (100%) | 38 (97.44%) | 39 (100%) |
| 25%(100) | 39 (100%) | 39 (100%) | 39 (10.%) | 39 (100%) | 39 (100%) | 39 (100%) | 39 (100%) | 39 (100%) | 38 (97.44%) | 39 (100%) |
| 28%(112) | 39 (100%) | 39 (100%) | 39 (100%) | 39 (100%) | 39 (100%) | 39 (100%) | 39 (100%) | 39 (100%) | 39 (100%) | 39 (100%) |

Table 6. Class Distribution of Arrhythmia data set

| Case | Class codes | Percentage of instances |
|---|---|---|
| **Commonly Occurring Classes (>=5%)** | 01,02,06,10,16 | 92% |
| **Rare Classes (<5%)** | 03,04,05,07,08,09,14,15 | 8% |

Table 7 Detected Rare Classes in Arrhythmia data set

| Number of Records | Number of Rare Classes Included (Coverage) | | | | | | | | | |
|---|---|---|---|---|---|---|---|---|---|---|
| | SOE1 ($\prod$) | SOE1 (+) | SOE1($S_q$) | | | SOE1 ($S_\infty$) | Find FPOF | Find CBLOF | Algorithm of [3] | KNN |
| | | | q=2 | q=5 | q=7 | | | | | |
| 85 | 33 | 32 | 33 | 34 | 33 | 11 | 32 | 32 | 43 | 28 |

From Table 7, we can see that the algorithm of [3] produced the best result with the cost of much more running time. In the remaining algorithms, most SOE1 variations are slight better, at least achieved the same level performance. Although the performance of SOE1 algorithm on identifying true outliers on this dataset is not so good as that of the algorithm of [3], but it is at least acceptable. And as we will show in next Section, SOE1 algorithm is very fast for larger dataset, which is more important in data mining applications.

## 5.5 Results on Synthetic Datasets

We present experiments with synthesized categorical data created with the software[2] developed by Dana Cristofor [37]. The data size (i.e., number of rows), the number of attributes and the number of classes are the major parameters in the synthesized categorical data generation. Table 8 shows the datasets generated with different parameters and used in the experiments. In all datasets, we set the random generator seed to 5.

Table 8. Test Synthetic Data Sets

| Data Set Name | Size | Number of Attributes | Number of Classes |
| --- | --- | --- | --- |
| DS1 | 100,000 | 10 | 10 |
| DS2 | 100,000 | 20 | 20 |
| DS3 | 100,000 | 30 | 30 |
| DS4 | 100,000 | 40 | 40 |

To demonstrate the scalability of our SOE1 algorithm, we choose Orca[3] algorithm [18], one of the fastest outlier detection algorithms known so far. We get the executable program of Orca from the author's website. Since the Orca software package comes with two programs, Orca and DPrep. Orca handles all of the computations associated with finding outliers. DPrep converts data sets that are stored as comma delimited text files into binary files for use with Orca. Hence, we first transform all datasets into binary files that can be used by Orca with Dprep. That is, we didn't include the preprocessing time in the running time of Orca. And the suggested default parameters are used in Orca.

Our SOE1 algorithm was implemented in Java. All experiments were conducted on a Pentium4-2.4G machine with 512 M of RAM and running Windows 2000.

Fig.3, Fig.4, Fig.5 and Fig.6 show the execution time of the two algorithms on four different dataset as the number of records is increased. As can be seen, the execution times of both algorithms increase with the increase of data size. However, the Orca algorithm dramatically increases its execution time with the increase of data size. In contrast, our algorithm always increases linearly.

In addition, SOE1 algorithm is always faster than Orca. Moreover, when data size is relatively large, SOE1 can be at least an order of magnitude faster than Orca.

Hence, we are confident to claim that SOE1 algorithm is suitable for mining very large dataset, which is very important in real data mining applications.

---

[2] The source codes are public available at: http://www.cs.umb.edu/~dana/GAClust/index.html
[3] The executable code of ORCA is public available at: http://newatlantis.isle.org/~sbay/software/orca/

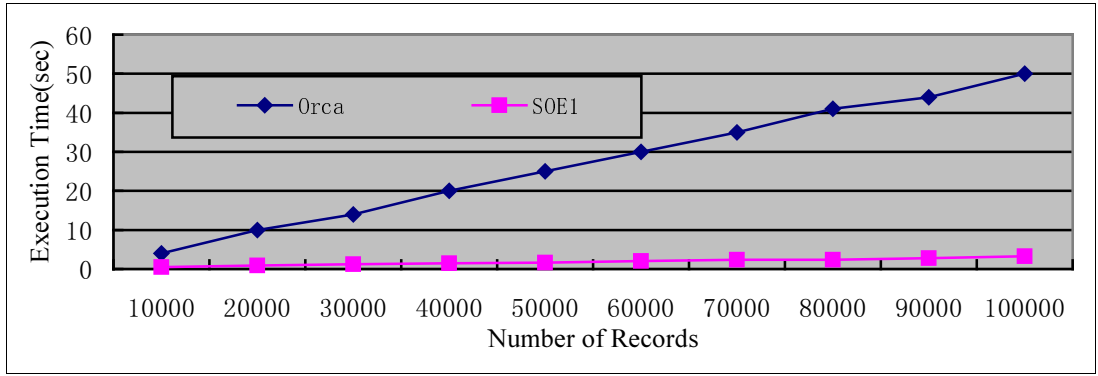

**Fig. 3.** The performance comparison between Orca and SOE1 on DS1

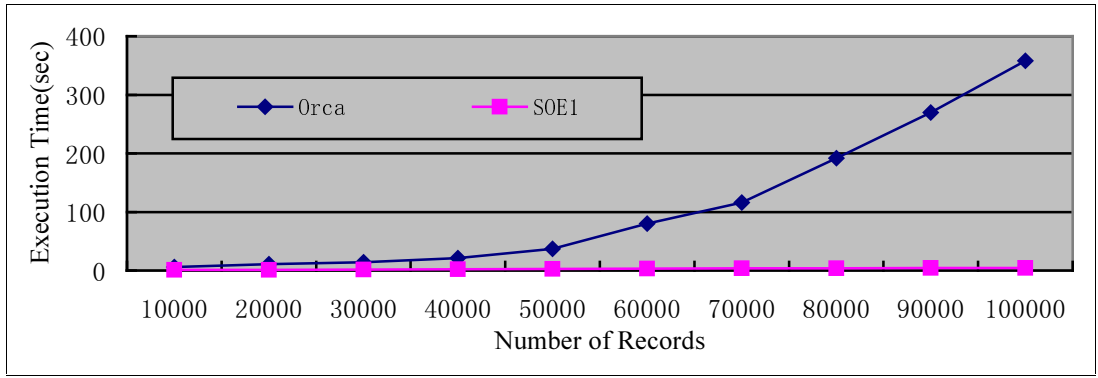

**Fig. 4.** The performance comparison between Orca and SOE1 on DS2

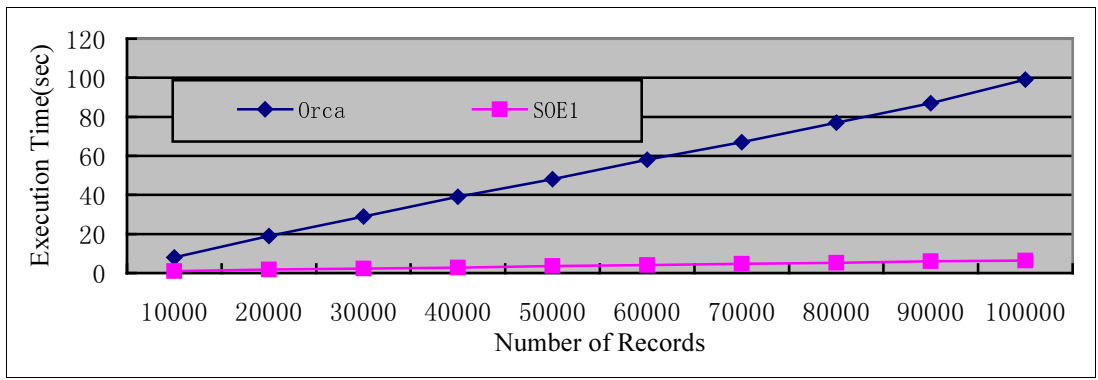

**Fig. 5.** The performance comparison between Orca and SOE1 on DS3

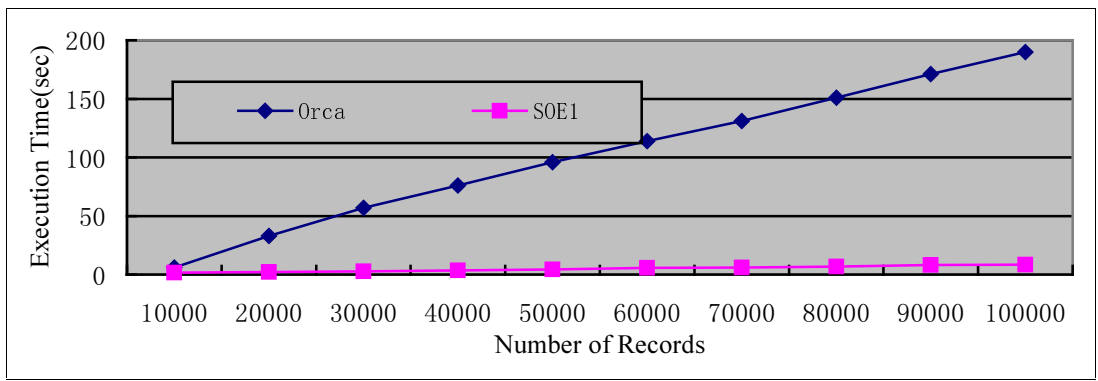

**Fig. 6.** The performance comparison between Orca and SOE1 on DS4

# 6 Conclusions

The main contributions of this paper can be summarized as follows:
- From the ensemble-learning viewpoint, a unified framework for outlier detection in high dimensional spaces is proposed.
- To our knowledge, it is the first work that uses ensemble-learning method in outlier detection research, i.e., propose the concept of outlier ensemble.
- Existing researches are unified in the proposed framework. Such integration may enable a better understanding of the problem of outlier detection in high dimensional spaces and help in devising improved or new algorithms.
- We developed a very simple and fast algorithm, called SOE1 in which only subspaces with one dimension is used for mining outliers from large datasets. Empirical study verified the superiority of SOE1
- The work done in this paper is no more than a preliminary exploration. Much further work is yet to be done in the future. For example, how to select only meaningful (at least meaningful from a statistical viewpoint) subspaces as input remains unaddressed. The general problem is open and provides promising future research directions.